\documentclass[prb,twocolumn,amsmath,notitlepage,showpacs]{revtex4-1}  
\usepackage{amsmath}
\usepackage{bm}
\usepackage{xcolor}
\usepackage{graphicx}

\usepackage{pdfsync} 

\usepackage{ifpdf}
\ifpdf
   \usepackage[unicode=true,colorlinks=true]{hyperref}
\fi

\begin{document}

\title{Anisotropic magneto-resistance in a GaMnAs-based
single impurity tunnel diode:\\ a tight binding approach}
\def\LPN{CNRS-Laboratoire de Photonique et de Nanostructures, 91460 Marcoussis, France}
\author{M.O.~Nestoklon}
\affiliation{Ioffe Physical-Technical Institute, Russian Academy of Sciences, St. Petersburg 194021, Russia}
\affiliation{\LPN}
\author{O.~Krebs}
\affiliation{\LPN}
\author{H.~Jaffr\`{e}s}
\affiliation{Unit\'e Mixte de Physique CNRS-Thal\`{e}s-Universit\'{e} Paris-Sud, 91767 Palaiseau Cedex, France}
\author{S.~Ruttala}
\affiliation{Unit\'e Mixte de Physique CNRS-Thal\`{e}s-Universit\'{e} Paris-Sud, 91767 Palaiseau Cedex, France}
\author{J.-M.~George}
\affiliation{Unit\'e Mixte de Physique CNRS-Thal\`{e}s-Universit\'{e} Paris-Sud, 91767 Palaiseau Cedex, France}
\author{J.-M.~Jancu}
\affiliation{FOTON-INSA Laboratory, UMR 6082 au CNRS, INSA de Rennes, 35043 Rennes Cedex, France}
\author{P.~Voisin}
\affiliation{\LPN}

\begin{abstract}
Using an advanced tight-binding approach, we estimate the anisotropy of the tunnel transmission associated with the rotation of the 5/2 spin of a single Mn atom forming an acceptor state in GaAs and located near an AlGaAs tunnel barrier. Significant anisotropies in both in-plane and out-of-plane geometries are found, resulting from the combination of the large spin-orbit coupling associated with the \textit{p-d} exchange interaction, cubic anisotropy of heavy-hole dispersion and the low $C_{2v}$ symmetry of the chemical bonds.
\end{abstract}


\maketitle

In the context of  semiconductor-based spintronics GaMnAs-based tunnel diodes are of utmost interest. In particular, it was recently demonstrated that the tunneling current in semiconductor heterostructures integrating the $p$-type ferromagnetic semiconductor \mbox{GaMnAs} is strongly affected by the direction of its magnetization, which can be arbitrarily changed  by an external magnetic field\cite{Gould_04, Ruster_05, Pappert_06, Jaffres_rev}. In this regard, two main phenomena must be distinguished: (i) the tunneling magnetoresistance (TMR) which is the dependence of the tunnel current with the respective parallel or antiparallel magnetic configurations of two ferromagnetic layers separated by a tunnel barrier giving rise to spin-valve effects and (ii) the tunneling anisotropic magnetoresistance (TAMR) which is the change of the tunnel current as a function of the magnetization direction of a (single) magnetic layer with respect to a reference coordinate system. So far, experimental results \cite{Gould_04, Ruster_05, Pappert_06} were obtained in the metallic regime, but it was suggested that the origin of TAMR could be still imputed to the anisotropic shape of the hole state bound to the isolated Mn impurity in a GaAs host\cite{Schmidt_07}. Various theoretical approaches indicate that the bound hole state is highly anisotropic\cite{Tang-Flatte_PRB05,MacDonald_09,Koenraad_11} and more extended in directions perpendicular to the Mn spin orientation. As a result, for in-plane magnetization, the exponential tail of hole bound state is more extended and tunneling probability higher. On the one hand, the validity of this explanation might be a matter of discussion for Mn densities close to the Mott limit (few percent Mn, considering a Bohr radius of 1 nm). On the other hand, even this qualitative picture needs to be checked by calculations including the effect of the barrier on the impurity state itself. Indeed, it is known from STM studies that hole wavefunction is strongly affected by its structural environment\cite{Jancu-STM,Krebs_09,Koenraad_11}. In this Letter we consider a model system close to experimentally achievable single impurity tunnel diode and estimate the TAMR effect in the framework of an advanced $sp^3d^5s^*$  tight-binding model\cite{Jancu98} including exchange interactions in the effective semi-classical approach\cite{Tang-Flatte_PRB05,MacDonald_09}. In addition to a strong dependency of tunnel probability on the angle between Mn spin orientation and the $[001]$ growth direction, our calculations reveal an unexpected in-plane anisotropy of comparable magnitude, with the $[110]$ and $[\bar110]$ eigenaxis characteristic of the $C_{2v}$ symmetry\cite{Krebs_96, Ivchenko_96}.

We model TAMR using the following scheme: we consider a single Mn impurity in a GaAs layer near a thin AlGaAs barrier (see Fig. 1) and calculate the neutral acceptor bound state. Tunneling transmission from this hole bound state through the AlGaAs barrier is estimated as an integral of the wavefunction tail in the GaAs "collector" layer on the right side of the barrier ($z>0$):
($\psi_h({\bm r})$) $\int_{z>0}\left| \psi_h({\bm r})\right|^2 d^3r$ . We do not consider here the "injector" problem, i.e. the way a hole is injected onto the impurity bound state from the left of the structure. This step can be ensured by p-doping the left electrode with shallow acceptors like Be, as was done in the context of STM studies\cite{Jancu-STM}.
\begin{figure}[h!]
  \includegraphics[width=0.8\linewidth]{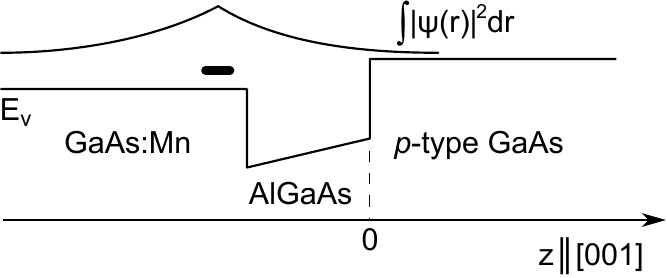}
  \caption{
     Schematic representation of a  Mn impurity in the GaAs layer near a thin Al$_{0.3}$Ga$_{0.7}$As barrier.
     Wavefunction of the hole bound to the acceptor penetrates this barrier. Tunneling probability
     is proportional to the integral of the hole wavefunction over the region to the right of the barrier.
  }\label{fig:scheme}
\end{figure}\\

For the physical validity of our calculations, all what matters is that the current be limited by the tunnel barrier in Fig. 1. The hole density of states (DOS) at the right collector like originally investigated by Bardeen \cite{Bardeen_61} is not explicitely included, as it just enters a prefactor that does not depend on Mn spin orientation. Still, this simple approach is only a zeroth order estimate of the tunnel probability as it does not take into account the spin structure of hole wave-functions. One may convince himself that this should be valid as soon as the kinetic energy at which the hole is injected into the collector is high enough (a few 10~meV's).  Indeed, in this case the admixture of heavy and light characters is such that the selectivity of transmission into the heavy- and ligh-hole bands should essentially vanish. A rough estimation shows that this corresponds to actual experimental situation. Note that a more complete treatment would require full description of a specific collector geometry.

To calculate wavefunctions we use the model and parameterization of Ref. \onlinecite{Jancu-STM} and include $p$-$d$ exchange interaction in the mean-field approach  previously introduced by Tang and Flatt\'e\cite{Tang-Flatte_PRB05}. This well established method, valid in the ferromagnetic regime, actually corresponds to an Ising coupling between the hole and Mn $3d^5$ electron spins. It amounts to treating the Mn 5/2 spin as a classical magnetic momentum\cite{MacDonald_09}.
In the paramagnetic regime a complete quantum treatment based on  Heisenberg exchange  coupling \cite{Bhattacharjee_99} would be definitely required, but it is much more difficult to implement self-consistently in the tight-binding formalism. It is also worth mentioning that for a given (classical) Mn spin orientation, the magnetic acceptor ground state is non-degenerate, in contrast with the four-fold degeneracy of a non-magnetic neutral acceptor state.

In practice, we use a supercell formalism, with a 8624 atom supercell of 7~ML$\times$7~ML$\times$22~ML, which is approximately 4~nm$\times$4~nm in lateral directions and 12~nm in growth direction. Technically, we destroy the effects of periodicity in the $[001]$ growth direction by adding an AlAs barrier to the left of the GaMnAs layer, that fully decouples adjacent cells. Periodic boundary conditions in lateral direction produce non-negligible effects on the bound state energy and on the tails of the wavefunction, that must be carefully characterized. However, these effects may hardly be considered as an artefact of the model as long as distance between Mn atoms corresponds to less than one Mn atom per thousand in the GaMnAs alloy. In real structures Mn atoms would be distributed randomly in the alloy while in the calculations they form a periodic array on a square lattice. We have checked that calculation results are stable with respect to Mn spatial distribution. Another point worth to be mentioned is the use of $sp^3d^5s^*$ model, compared to the $sp^3$ model used in previous papers\cite{Tang-Flatte_PRB05, MacDonald_09}. While $sp^3$ is a qualitatively acceptable basis to discuss valence band properties in III-V compounds, it is well-known\cite{Boykin_97} that it fails to reproduce accurately the Luttinger parameters that describe valence band dispersion. $sp^3d^5s^*$ is actually the minimal tight-binding model which allows fitting all the parameters of valence band over a large energy range, and in particular, the cubic anisotropy (warping) of the heavy-hole dispersion. This is essential for an accurate account of the kinetic energy part of the impurity hamiltonian. In order to image the wavefunctions, the local density of states at atomic sites is approximated by gaussian functions with a radius depending on orbital type ($\sim$1.5-2{\AA}).

\begin{figure}
  \includegraphics[width=\linewidth]{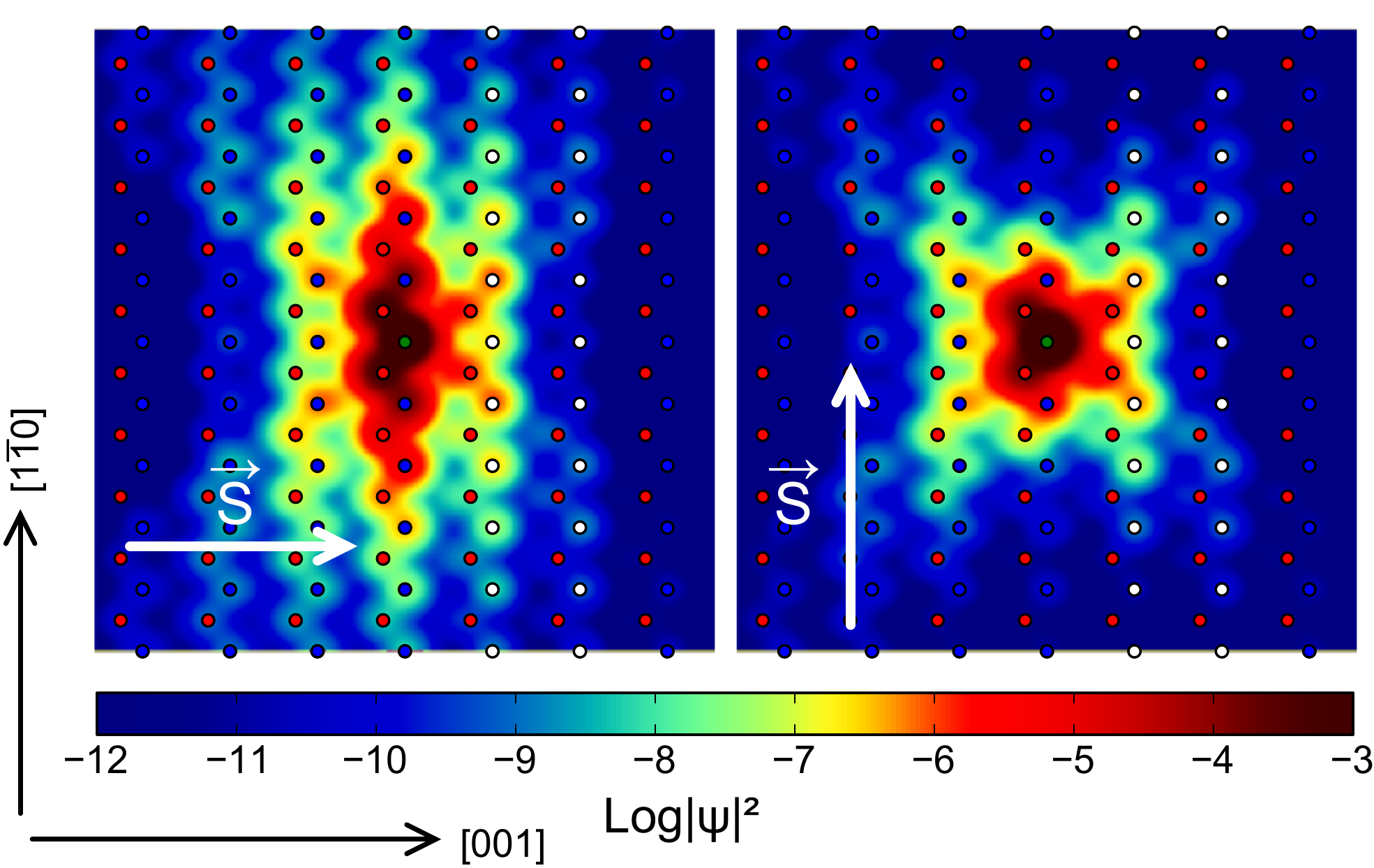}
  \caption{(Color online)     Cross-section of the neutral acceptor wavefunction in the (110) plane containing the Mn
     dopant. Wavefunction amplitude is color-coded according to the log scale. Atomic
     positions are shown with circles~: red and blue for As and Ga, green for Mn, and white
     for the virtual Al$_{0.3}$Ga$_{0.7}$ cation in the tunnel barrier.
  }\label{fig:wavefunctions}
\end{figure}

The cross-sections of the wavefunction of a hole bound to the Mn acceptor are shown in Fig.~2 for two directions of the Mn spin, in the $(110)$-plane containing the impurity. The wavefunction amplitude is color-coded according to the log scale below the panels, and atomic positions are indicated by colored dots (see figure caption). Left panel displays the situation for a Mn spin along the $[001]$ direction, and right panel for a Mn spin in $[1\bar10]$ direction. While it is not obvious from these figures how tunneling probability can be affected, it may be concluded that the hole wavefunction changes significantly in size as already emphasized in Ref.~\onlinecite{Tang-Flatte_PRB05}. The effect of barrier proximity is clearly visible from the left/right asymmetry of wavefunction tails, and the role of lateral periodicity is evident from the non vanishing tails in the vertical direction in Fig.~2 (supercell in the calculations is few ML larger in lateral direction than shown in figure). Neutral acceptor binding energy decreases when the impurity gets closer to the barrier\cite{Bastard}. For Mn located at 1 ML from the barrier, this reduction amounts to 20 meV. The eigenenergies also change with Mn spin orientation: this is associated with the magnetization-dependence of the wavefunction tails, and the related change in tunnel coupling of acceptors in adjacent cells\cite{MacDonald_09} . The amplitude of this change for Mn located at 1~ML from the barrier as a function of angle between Mn spin and growth direction is almost 10~meV. Under in-plane rotation of the spin direction this modulation of eigenenergies is reduced but still amounts to about 4~meV. The calculated tunneling probability is extremely sensitive to the impurity distance from the barrier, decreasing exponentially with characteristic length $\sim$3.7~{\AA}. However, for all distances it shows similar angle dependencies. In the following, we focus on the 1~ML distance corresponding to Fig.~2.

\begin{figure}
  \includegraphics[width=0.95\linewidth]{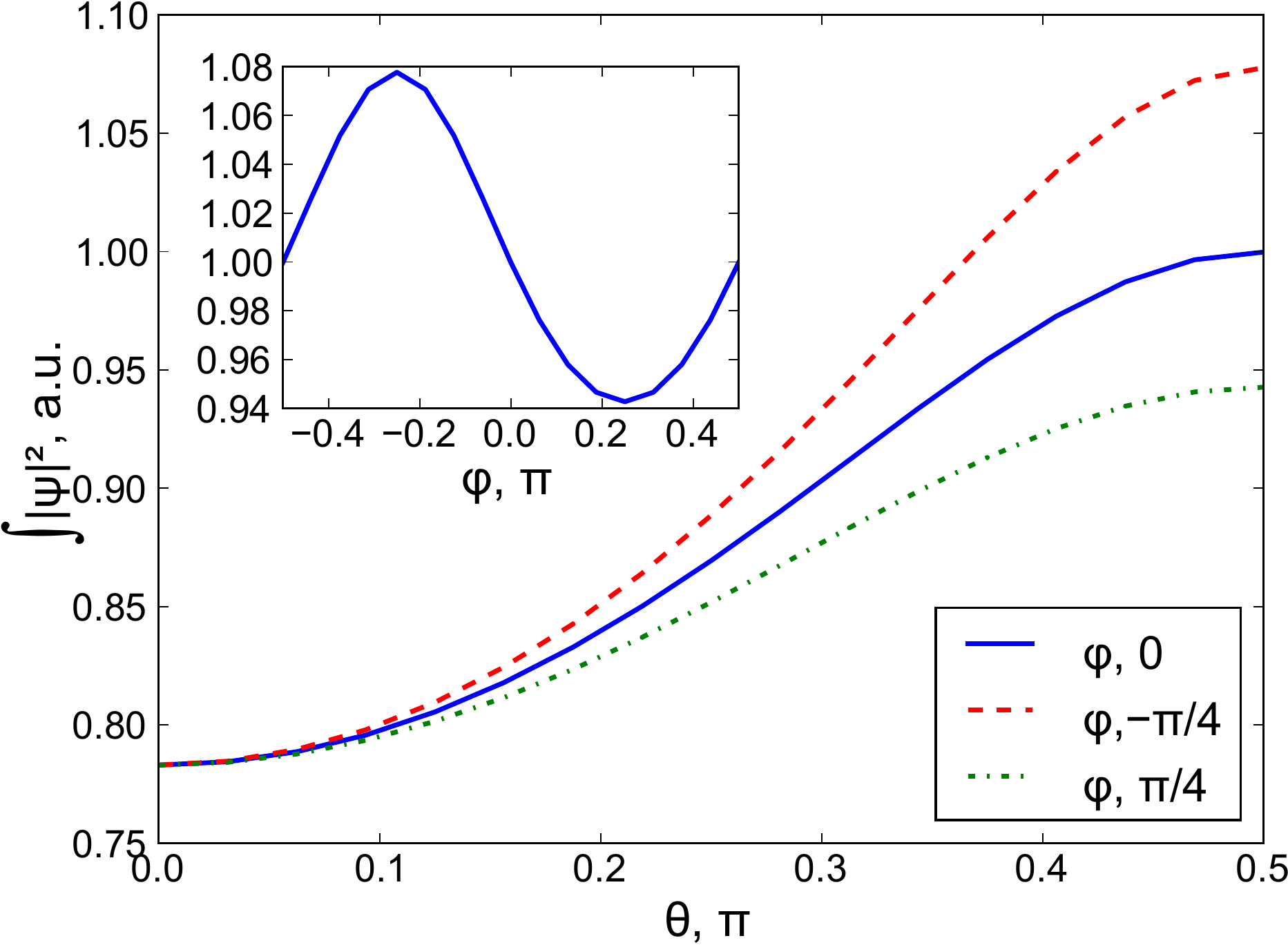}
  \caption{
     Dependence of wavefunction integral on right side of barrier on Mn spin orientation.
     a) dependence on zenithal angle $\theta$ for different values of the azimuthal angle $\varphi$ with
     respect to [100], b) azimuthal plot of in-plane anisotropy.
  }\label{fig:theta}
\end{figure}

Calculation results are illustrated in Fig.~3. Mn spin direction is defined with the zenithal angle $\theta$ with respect to $[001]$ growth direction and in-plane azimuthal angle $\varphi$ with respect to $[100]$ axis. The dependency on $\theta$ follows the qualitative explanation given in the introduction: the wavefunction of the hole localized at acceptor has a somewhat oblate shape and follows Mn spin direction. If the spin is oriented along the growth direction, the barrier lies in the direction where the exponential decay is stronger, giving a lower tunneling probability by more than 20\%, which was defined in a previous letter as a negative TAMR~\cite{Elsen_07}. Wavefunction deformation due to the barrier proximity and large cubic anisotropy due to valence band warping (see Fig.~2) do not change this qualitative argumentation significantly. In-plane variation (inset in Fig.~3) is more intriguing. It has similar amplitude as the zenithal variation, and shows the characteristic $C_{2v}$ symmetry, with pronounced differences between $[\bar110]$ and $[110]$ directions. We have checked that this result is not a calculation artifact due to lateral periodic boundary conditions by rotating the supercell orientation, which is equivalent to rotating the square lattice of Mn atoms: this rotation has no influence on the calculated tunnel current anisotropy. This suggests that these results would be qualitatively unchanged for the randomly positioned Mn atoms. In-plane anisotropy is also nearly unaffected by the lateral supercell size, although the variation of bound-state eigenenergy with spin orientation increases with decreasing size (or, equivalently, with increasing Mn concentration). The calculated in-plane anisotropy of about 14\% actually results from a combination of three factors reducing the symmetry of bound state wavefunction~: i) a native anisotropy of the bulk wavefunction, which is clearly evidenced by removing the barriers in the supercell while calculating the integral over the same part of the wavefunction~; ii) the left/right asymmetry of the wavefunction due to the interface proximity~; and iii) the interface $C_{2v}$ symmetry itself\cite{Krebs_96, Ivchenko_96}. These contributions interfere and can hardly be separated in the final result. \\
\indent In summary, we have estimated  the anisotropy of the tunnel transmission from the fundamental A$^0$ hole state attached to a single Mn dopant in a GaAs host matrix, coupled to a reservoir through an AlGaAs tunnel barrier. The simulated structure displays pronounced anisotropy of the tunnel transmission as a function of Mn classical spin direction. Significant in-plane anisotropy of the tunnel current is obtained. Yet, a full treatment of Heisenberg (instead of Ising) spin coupling and direct account of applied magnetic field would be highly desirable in order to fully assert the potential of single impurity TAMR effect.

Acknowledgment: This work was supported by ``Triangle de la Physique'', by CNRS-RAS joint laboratory ILNACS and by the European Networks for Initial Training (ITN): SemiSpinNet. 
\bibliography{TAMR}
\end{document}